\documentclass{emulateapj}

\usepackage{epstopdf}
\usepackage{color}

\newcommand{\myemail}{abik@mpia.de}

\newcommand{\brg}{Br$\gamma$}
\newcommand{\HII}{H{\sc ii}}

\defcitealias{Indebetouw05}{Ind05}
\defcitealias{Nishiyama09}{NI09}

\shorttitle{Age spread in W3 Main}
\shortauthors{Bik et al.}

\begin{document}


\title{Age spread in W3 Main:  LBT/LUCI near-infrared spectroscopy of the massive stellar content\altaffilmark{1}}


\author{A. Bik\altaffilmark{2}, Th. Henning\altaffilmark{2}, A. Stolte\altaffilmark{3}, W. Brandner\altaffilmark{2}, D. A. Gouliermis\altaffilmark{2}, M. Gennaro\altaffilmark{2},  A. Pasquali\altaffilmark{4}, B. Rochau\altaffilmark{2}, H. Beuther\altaffilmark{2},  N. Ageorges\altaffilmark{5}, W. Seifert\altaffilmark{6}, Y. Wang\altaffilmark{7}, N. Kudryavtseva\altaffilmark{2}}
\email{\myemail}

\altaffiltext{1}{Based on data acquired using the Large Binocular Telescope (LBT). The LBT is an international collaboration among institutions in Germany, Italy and the United States. LBT Corporation partners
are: LBT Beteiligungsgesellschaft, Germany, representing the MaxPlanck Society, the Astrophysical Institute Potsdam, and Heidelberg University; Istituto Nazionale di AstroÞsica, Italy; The University of
Arizona on behalf of the Arizona university system; The Ohio State
University, and The Research Corporation, on behalf of The University
of Notre Dame, University of Minnesota and University of Virginia.}
\altaffiltext{2}{Max-Planck-Institut f\"ur Astronomie, K\"onigstuhl 17, 69117 Heidelberg, Germany}
\altaffiltext{3}{Argelander Institut f\"ur Astronomie, Auf dem H\"ugel 71, 53121 Bonn, Germany}
\altaffiltext{4}{Astronomisches Rechen Institut, M\"onchhofstrasse 12 - 14, 69120 Heidelberg, Germany}
\altaffiltext{5}{Max-Planck-Institut f\"ur extraterrestrische Physik, Giessenbachstrasse 1, 85748 Garching, Germany}
\altaffiltext{6}{Landessternwarte K\"onigstuhl, Zentrum f\"ur Astronomie Heidelberg, K\"onigstuhl 12, 69117 Heidelberg, Germany}
\altaffiltext{7}{Purple Mountain Observatory, Chinese Academy of Sciences, 210008, Nanjing, PR China}

\begin{abstract}
We present near-infrared multi-object spectroscopy and $JHK_{\rm{s}}$ imaging of the massive stellar content of
the Galactic star-forming region W3 Main, obtained with LUCI  at the Large Binocular Telescope. We confirm 15  OB stars in W3 Main and derive spectral types between O5V and B4V from their absorption line spectra. Three massive Young Stellar Objects are  identified by their emission line spectra and near-infrared excess. 
The color-color diagram of the detected sources allows a detailed investigation of the slope of the near-infrared extinction law towards W3 Main. Analysis of the Hertzsprung Russell diagram  suggests that the Nishiyama extinction law fits the stellar population of W3 Main best ($E(J-H)/E(H-K_{\rm{s}}$) = 1.76 and $R_{\rm{K_s}}$ = 1.44).  From our spectrophotometric analysis of the massive stars and the nature of their surrounding \HII\ regions we derive the evolutionary sequence of W3 Main and we find evidence of an age spread of at least 2-3 Myr. While the most massive star (IRS2) is already evolved, indications for high-mass pre--main-sequence evolution is found for another star (IRS N1), deeply embedded in an ultra compact \HII\ region, in line with the different evolutionary phases observed in the corresponding \HII\ regions. We derive a  stellar mass of W3 Main of ($4 \pm 1) \times 10^3 M_{\sun}$, by extrapolating from the number of OB stars using a Kroupa IMF and correcting for our spectroscopic incompleteness.  We have detected the photospheres of OB stars from the more evolved diffuse \HII\ region to the much younger UC\HII\ regions, suggesting that these stars have finished their formation and cleared away their  circumstellar disks very fast. Only in the hyper-compact \HII\ region (IRS5), the early type stars seem to be still surrounded by circumstellar material.
\end{abstract}


\section{Introduction}
Despite the impact on their surroundings, the formation and early evolution of massive stars is poorly constrained, primarily because of their scarcity and short lifetimes.   OB stars are usually observed in very young star-forming regions, e.g. clusters and associations, which are still embedded. Near-infrared observations are, in most cases, the only way to study the stellar content of these young regions. However, a pure photometric characterization of young embedded stellar clusters is strongly hampered by highly varying extinction, unknown distances and infrared excess of the young cluster members.

On the other hand, spectroscopy offers an unambiguous identification of the massive stellar content and has proven to be a powerful method to find and characterize the newborn OB stars. Stellar properties, like effective temperature and luminosity, are derived based on the spectral features and extinction, distance and possible infrared excess can be determined reliably \citep[e.g.][]{WatsonSpectra97, Blum00, Hanson02,Ostarspec05,Puga06,Bik10,Puga10}. By comparing the effective temperature and  luminosity to  stellar isochrones, the age as well as the mass of the stellar population in the young cluster can be derived more reliably  than by photometry alone \citep{Bik10}, as the large extinction variations  affects  the identification of a clearly reddened cluster main sequence.  Apart from the basic properties of the star clusters, also age spread inside the cluster, the evolution of circumstellar disks and the early evolution of the young OB stars can be characterized and studied.

Analyses of increasing samples of stars in the same star cluster have produced evidence for age spreads in several cases. Based on the massive star content, \citet{Clark05,Negueruela10} find very little age spread (less than 1 Myr) in starburst cluster Westerlund 1. However, in the Orion Nebula Cluster age spread has been found based on pre-main-sequence (PMS) stars \citep{daRioOrion10}. \citet{Kraus09} presented evidence for an age spread in Taurus.  Additional evidence for an age spread has been found in S255 where the most massive star is still deeply embedded and driving an outflow, while the PMS population has an age of 1-3 Myr \citep{Wang11}. \citet{Kumar06} found evidence for stellar clusters around high-mass protostellar objects, suggesting a similar age spread.

Additionally, young stellar clusters are the places to look for signatures of how massive stars have formed. Remnant accretion disks are being dispersed on a short timescale \citep{Hollenbach94}. However, massive Young Stellar Objects (YSOs) have been identified in several high-mass star forming regions \citep[e.g.][]{Blum04,Brgspec06}.  Studying the frequency of massive YSOs as function of cluster age and mass will allow us to place time scales on the disk evaporation process.

Before arriving on the main sequence,
low- and intermediate mass  stars are still contracting and are cooler and bigger than they will be on the main sequence. The intermediate-mass stars are detected as G and K-type stars and will evolve to A and F stars respectively \citep{Bik10}.  Theoretical predictions show that stars more massive than 10 $M_{\sun}$ do not have an optically visible PMS phase \citep{Palla90}.  However, using deep near-infrared observations  we can probe objects  deeply embedded behind up to 50 magnitudes of visual extinction and we might follow  the PMS phase of the massive stars.

Theoretical modeling of  high-mass  protostellar evolution predicts that massive  protostars might migrate towards the main sequence following a path similar to intermediate mass PMS stars \citep{Yorke08,Hosokawa09,Hosokawa10}. Due to the high accretion rates, the massive protostars will swell up to a maximum radius of 125 $R_{\sun}$ and exhibit a relatively cool temperature of $\sim$5000 K. In the HRD they will look like red supergiants \citep{Morino98,Linz09}.
After most of the stellar mass is acquired, the protostar is contracting to the main sequence and will gradually reach its main sequence size and temperature \citep{Zinnecker07}. 

In this paper, we address these issues of massive star formation with near-infrared $K$-band multi-object spectroscopy as well as $JHK_s$ imaging of the massive stellar content of W3 Main, obtained with LUCI at the LBT.  We perform for the first time a spectral classification of its massive stellar content, allowing the detailed assessment of the evolutionary status of these stars and their \HII\ regions. 
The W3  \citep{Westerhout58}  region is part of an extended star formation complex in the Perseus spiral arm and is  associated with W4 and W5 \citep[see][for an exhaustive review]{Megeath08}, spanning an area of 200 $\times$ 70 pc. W3 is located at a distance of 1.95 $\pm$ 0.04 kpc \citep{XuW306}. 
The W3 molecular cloud has a shell  morphology wrapped around two optically bright \HII\ regions; IC 1795 and NGC  896 of which IC 1795 has an age of 3-5 Myr \citep{OeyW305,Roccatagliata11}. A much younger part of W3 is still deeply embedded in the molecular cloud surrounding the OB association and consists of 3 embedded clusters of compact \HII\ regions: W3-North, W3-OH and W3 Main. The latter region is the subject of this paper.

Radio continuum observations  reveal several compact and diffuse \HII\ regions in the W3 Main area  \citep{WynnWilliams71,Harris76} as well as ultra-compact \HII\ (UC\HII)  and hyper-compact \HII\ (HC\HII) regions  \citep{Claussen94,Tieftrunk97}.  The location of the C$^{18}$O \citep{Tieftrunk95} and NH$_{3}$ \citep{Tieftrunk98} molecular emission peaks in a radio quiet area south of the region W3 D. Sub-mm dust continuum emission is found here too, as well as centered on IRS5 \citep{Moore07}.

Along with the first detection of the radio continuum sources, the first near-infrared imaging revealed several bright point sources associated with these \HII\ regions \citep{WynnWilliams72}. More sensitive imaging surveys \citep{Hayward89,Tieftrunk98,Ojha04} have revealed the stellar richness of the W3 Main region and identified all the candidate ionizing sources of the \HII\ regions via photometry.  In the X-rays, where the extinction is much less severe, the entire  PMS population is detected, and the cluster shows a spherical geometry with a size of 7 pc, about twice as large as seen in the near-infrared \citep{Feigelson08}.

The different evolutionary phases of the \HII\ regions make W3 Main an ideal target to study age spread and the evolution of circumstellar disks around massive stars. The paper is organized as follows; in Section 2 we discuss the LUCI imaging and spectroscopic observations and data reduction. Section 3 presents the near-infrared photometry, the spectroscopic classification and the derived HRD. In section 4 we discuss the evolutionary stage of the massive stars as well as the detected age spread,  and summarize the conclusions in Section 5.

\section{Observations and data reduction}

Near-infrared observations of W3 Main were obtained with LUCI 1 (LBT NIR spectroscopic Utility with Camera and Integral-Field Unit for Extragalactic Research),  mounted at the Gregorian focus of the Large Binocular Telescope (LBT) on Mount Graham, Arizona \citep{Hill06}. LUCI 1 \citep{Ageorges10,Seifert10} is  a near-infrared multimode instrument operating in seeing-limited imaging, long-slit spectroscopy as well as multi-object spectroscopy (MOS) mode \citep{Buschkamp10}.

\begin{deluxetable*}{lrrrl}
\tablewidth{0pt}
\tablecaption{Observing details \label{tab:log}}
\tablehead{
\colhead{Observations} &
\colhead{Date (UT)} &
\colhead{DIT(s)\tablenotemark{a}} &
\colhead{Exp time (s)} &
\colhead{Objects} 
}
\startdata
$J$   & 2009/12/18 &  5   & 800 & \\
$H$  & 2009/12/19 &  5   & 1000 & \\
$K_{\rm{s}}$  & 2009/02/05 &  5   & 900 & \\
W3Klt14K		& 2009/12/19 &  60 & 1800 & IRS N3				\\
W3Klt14Km2	& 2009/12/18 &  60 & 1800 & IRS3a, IRS5, IRS N1, IRS N6	\\
LSS			& 2011/03/06 & 60 & 2400 & IRS4, IRS7	\\
W3BrightK	& 2009/12/17 & 30 & 1080 & other objects
\enddata
\tablenotetext{a}{Detector Integration Time}
\end{deluxetable*}

\subsection{Imaging observations}
W3 Main at $\alpha$(2000) = 02$^h$25$^m$37.5$^s$,  $\delta$(2000) = +62$^{\circ}$05\arcmin 24\arcsec, was imaged in $JHK_{\rm{s}}$ with LUCI. The $J$ and $H$ data were taken on December 18 and December 19, 2009, respectively, while the $K_{\rm{s}}$ data were obtained on Feb 5, 2010  with the N3.75 camera, providing a 4\arcmin$\times$4\arcmin\ field of view at an image scale of 0.12\arcsec /pixel.
 
The imaging data were taken with a detector integration time (DIT) of 5 sec and a NDIT (number of integrations) of 10.  A random dither pattern has been applied using  8 up to a maximum of 10 different positions and a jitter box of 30\arcsec. The total exposure times per filter are listed in Table \ref{tab:log}. Sky frames were obtained using the same instrument settings immediately afterwards, centered on a off-cluster field at  $\alpha$(2000) = 02$^h$25$^m$41.0$^s$,  $\delta$(2000) = +62$^{\circ}$12\arcmin 32\arcsec.  The  final image quality of the  of the $J$ and $H$ images is 0.6\arcsec and 0.9\arcsec\ for the $K_{\rm{s}}$ image.

\subsection{Spectroscopic observations}
W3 Main was observed with the LUCI MOS unit in the $K$-band on December 17 - 19, 2009. To ensure a proper alignment of the masks for the MOS observations, $K$-band pre-imaging was acquired on November 01, 2009. 
Three masks were created to cover the stellar content ranging from the most massive stars to the intermediate-mass PMS stars.  We designed the masks such that the difference in magnitude between the stars in the mask was around 1 magnitude, ensuring comparable signal-to-noise ratio for all spectra in one mask. 

We used the 210$\_$zJHK grating in combination with a slit width of 1.0\arcsec, providing us with a resolution $R$=4000. 

The spectra were collected with a spatial resolution of 0.25\arcsec/pixel (N1.8 camera), providing the largest wavelength coverage ($\Delta \lambda$ = 0.328 \micron). 
The  MOS slits were typically 4\arcsec\ long, to allow for a 2\arcsec\ nodding for sky subtraction while minimizing the contamination by other stars.

In addition to the MOS spectra, we obtained a long slit spectrum of IRS4 and IRS7 on March 6, 2011 with a total integration time of 2400 sec. The instrument setup was the same as for the MOS spectroscopy. For the long-slit observations, the nodding offset  was set to 45\arcsec.  The details of the observations are shown in  Table \ref{tab:log}, which lists the different DITs and total exposure times used for the masks and long-slit spectra.  The seeing of the spectroscopy data  varied between 0.6\arcsec and 0.9\arcsec. 

Telluric standard stars of early B type were observed immediately before and after the science frames. The standard star was observed using the same mask as the science observations, as well as being placed in two different slits,  covering the blue and the red part of the spectral band respectively. Immediately after the science and standard star observations, arc lamp and flat fields were taken  to minimize the effects of flexure.

\subsection{Data reduction}

\subsubsection{Imaging}

\begin{figure*}
\plotone{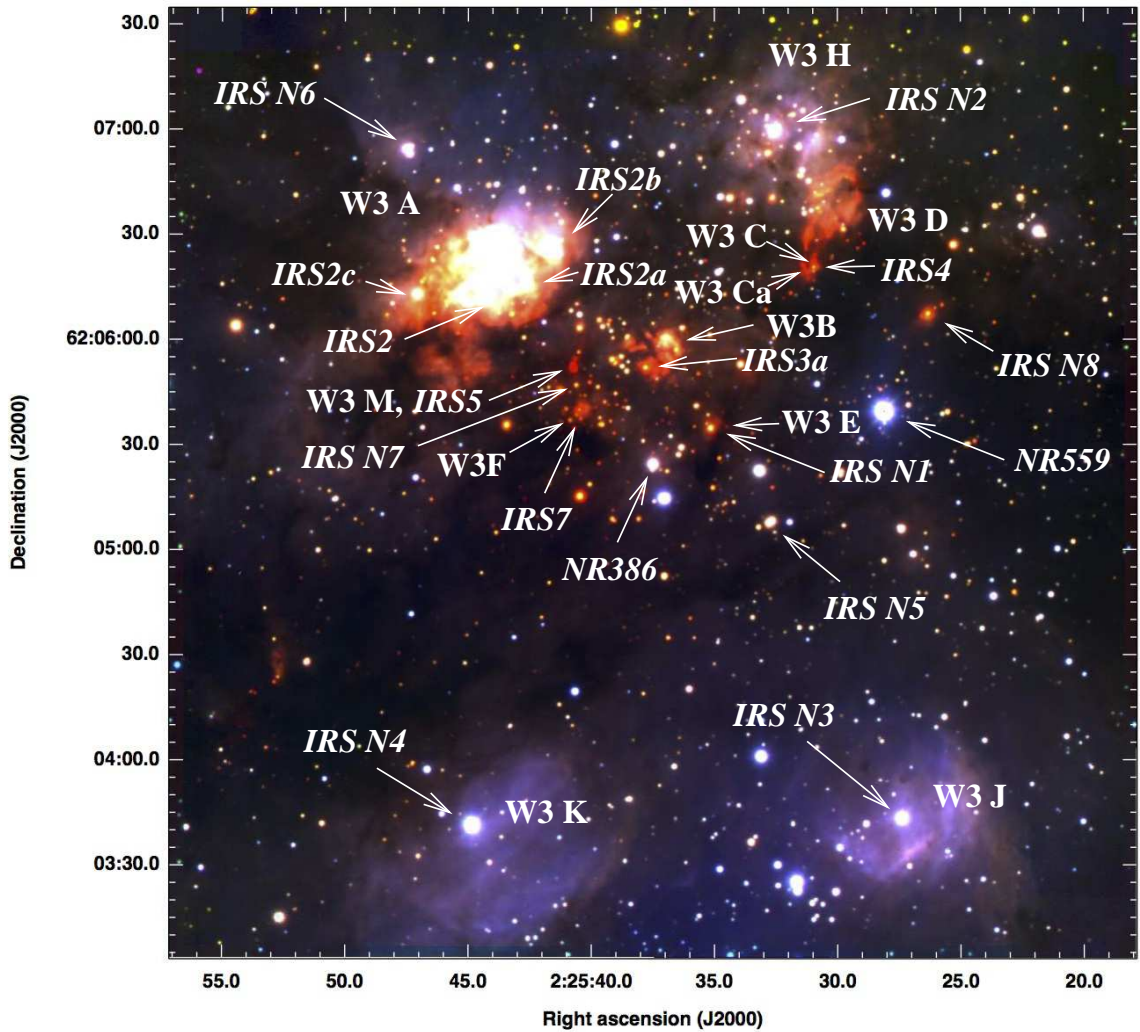}
\caption{LUCI $JHK_{\rm{s}}$  color image of W3 Main, North is up and East is left. The observed field of view is 4.5\arcmin $\times$4.5\arcmin, corresponding to 2.55$\times$2.55 pc at a distance of 1.95 kpc. The brightest region of extended emission in the north-east is the \HII\ region W3 A. The two diffuse, spherical \HII\ regions in the south are W3 K (south-east) and W3 J (south-west). In the center, the deeply embedded massive YSO IRS5 can be found. Annotated are the \HII\ regions \citep{Tieftrunk97} as well as the candidate massive stars of which the spectra are presented in this paper (in italic). Additionally, the location of the 2 massive stars (NR386, NR559) identified by \citet{Navarete11} are identified. \label{fig:W3JHK} } 
\end{figure*}

The near-infrared $JHK_{\rm{s}}$ images were reduced with standard IRAF\footnote{IRAF is distributed by the National Optical Astronomy Observatory, which is operated by the Association of Universities for Research in Astronomy, Inc., under cooperative agreement with the National Science Foundation.} routines, as described in \citet{Pasquali11}. The images where corrected for dark current and flat fielded using sky flats taken during morning and evening twilight. A  sky frame was created by combining the images taken at the offset position using a weighted mean and rejecting the lowest 3 and highest 8 values to remove the stars from the frames.
The master sky frame was subtracted from the individual science frames.  After background subtraction, the images were corrected for geometric distortion. An astrometric solution was derived using stars in common with 2MASS \citep{Skrutskie06}.  The corrected frames where combined and weighted by the average level of the background, producing the final images (Fig. \ref{fig:W3JHK}).

Photometry on the $JHK_{\rm{s}}$ images was performed using the {\em starfinder} software  \citep{Diolaiti00}.  This package is designed  to fit an empirical point spread function (PSF) to a crowded stellar field. The stellar PSF was constructed using 22 bright and isolated sources in the frame.  The radial extent of the PSF was set to 40 pixels,  beyond the radius at which the PSF structure was dominated by background noise.  The final PSF  was created iteratively after selection and subtraction of neighboring sources from the image. Three iterations were performed before the final PSF was extracted. 
All sources  at $\geq$ 3 $\sigma$ above the local background noise
were included in the detected source list, and three PSF fitting
iterations were performed to extract the faintest sources.

We verified the results of our {\em starfinder} photometry by performing an additional run with the {\sc daophot} package \citep{Stetson87} within the IRAF environment. {\sc daophot} treats the background and the PSF model differently than {\em starfinder}, allowing us thus to check our photometric results from different approaches. Stellar sources were detected with the {\em daofind} task and aperture photometry was performed with the {\em phot} task in radii $\sim$ 1 to 2~$\times$ the FWHM of the PSF. For each filter a reference PSF was constructed by combining the PSF of at least 20 objects with the tasks {\em pstselect} and {\em psf}. PSF-fitting photometry was performed with the {\em allstar} task, using the PSF models to fit all objects identiÞed with a 3$\sigma$ confidence level over the local background. Our results from both photometric runs are very similar, however, {\em starfinder} is able to better treat the saturated stars in the images.

Several of the bright stars in the \HII\ region appeared saturated in our $JHK_{\rm{s}}$ images. The $J$-band image has 7 stars with counts above the linearity limit, while the $H$ and $K_{\rm{s}}$  images have 8 stars in the non-linear regime of the detector.  To obtain reliable magnitudes for these bright stars, {\em starfinder}  was used to fit the wings of the saturated stars with the determined PSF to reconstruct the flux of the saturated part.  However, a few of the brightest stars could not be recovered by {\em starfinder} due to their heavily saturated cores having negative counts. For those stars, the negative core of the PSF was removed and the peak of the PSF was flattened.  The same area was also flattened in the reference PSF, and the best fit determined, such that the differences in the linear part of the star's and the reference PSF were minimized. 

This procedure is compared to the recovery done by \emph{starfinder} on the non-saturated stars allowing the determination of the uncertainties of the used method. The measured uncertainties where  0.1 mag for $J$, 0.07 mag for the $H$-band and 0.08 mag for $K_{\rm{s}}$. Comparison between the obtained magnitudes and the 2MASS values for the isolated stars shows a match within 0.1 mag for all three bands, suggesting that the above described method delivers reliable values. 

Finally, we cross-matched the obtained catalogs for each filter to identify the sources that are detected in more than one  band. We allowed a maximum offset of 4 pixels  (0.48\arcsec)  for the center of each point source.  We calibrated the  LUCI photometry with  2MASS \citep{Skrutskie06}.
The matching radius was 4 pixels after transforming the
2MASS positions to the LUCI x,y coordinate system. Only stars with
magnitudes $10 < J < 16.5$ mag, $10 < H < 14.5$ mag, and 9.5 $< K_{\rm{s}} <$ 14.0 mag
were used to avoid saturation effects in LUCI and ensure reliable
photometry in 2MASS at the faint end. As mismatches in data sets with
different spatial resolutions and sensitivities are frequent, we
selected stars with $|$mag(LUCI) - mag(2MASS)$|$ $\leq$ 0.5 mag from the
median zero point of all stars in each filter. This led to a selection
of 54 stars in $J$, 49 stars in $H$ and 53 stars in $K_{\rm{s}}$  as final calibrators,
yielding the zero points $J_{\rm{zpt}}$ = 29.205 $\pm$ 0.022 mag, $H_{\rm{zpt}}$ = 29.072 $\pm$ 0.014 mag,
$K_{\rm{s,zpt}}$ = 28.373 $\pm$ 0.018 mag.

To obtain a realistic estimate of the photometric errors, we divided the datasets in two subsets with identical total exposure time and created two auxiliary images for each filter.  An empirical psf was constructed in the same way as on the deep dataset and  \emph{starfinder} photometry was derived for the auxiliary datasets. The photometric uncertainties are derived from the difference in magnitude between the auxiliary datasets. For faint sources, not detected in the auxiliary datasets, the largest error in the same magnitude range, as derived for other sources, was applied as a conservative error estimate.

The final photometric catalog, containing all the sources detected in the 3 bands, has a limiting magnitude of $\sim$21.5 mag in $J$, $\sim$19.5 mag in $H$ and $\sim$19.0 mag in $K_{\rm{s}}$. A more detailed assessment of the completeness limits will be done in a future paper where the entire stellar content of W3 Main will be discussed.

\begin{figure*}
\plottwo{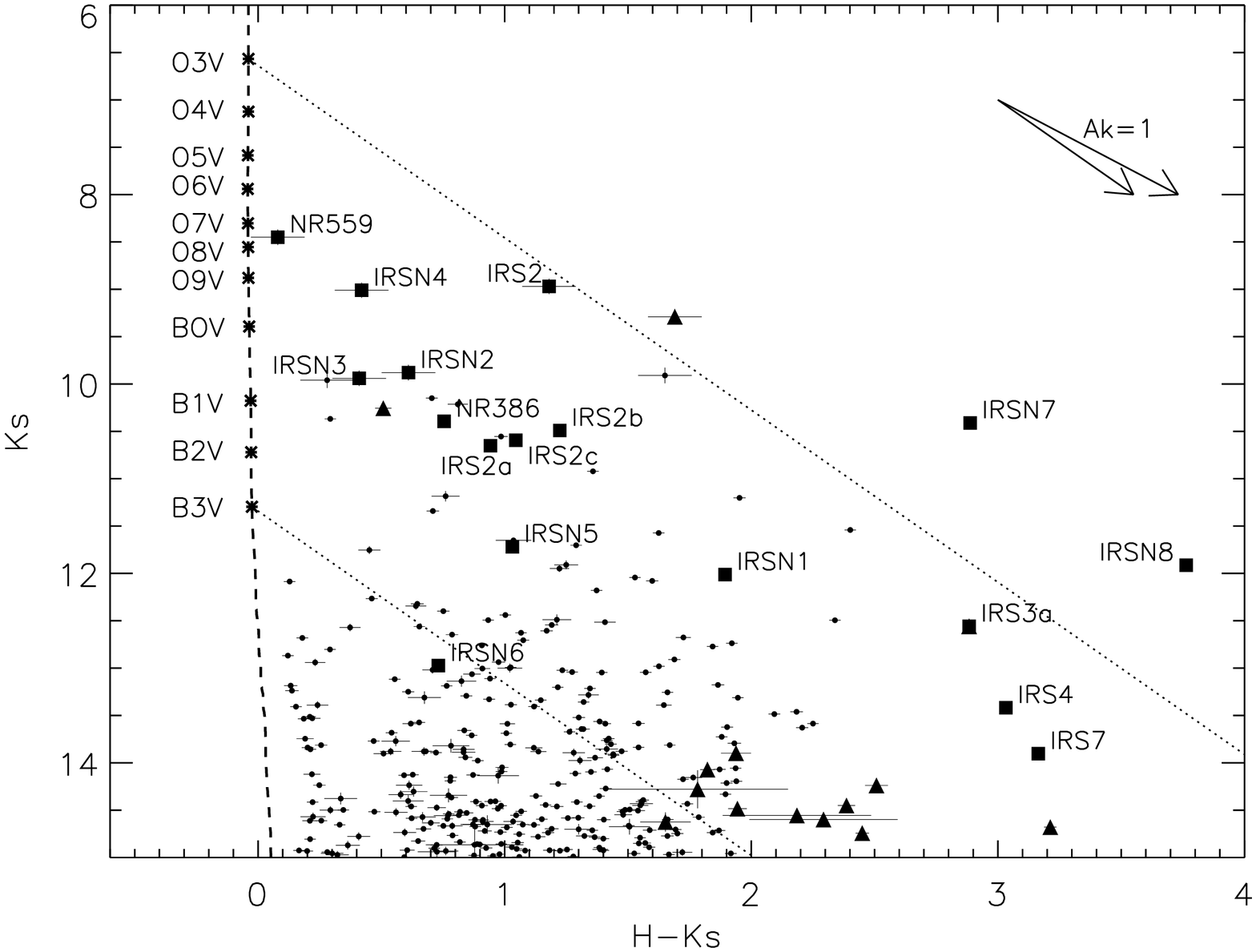}{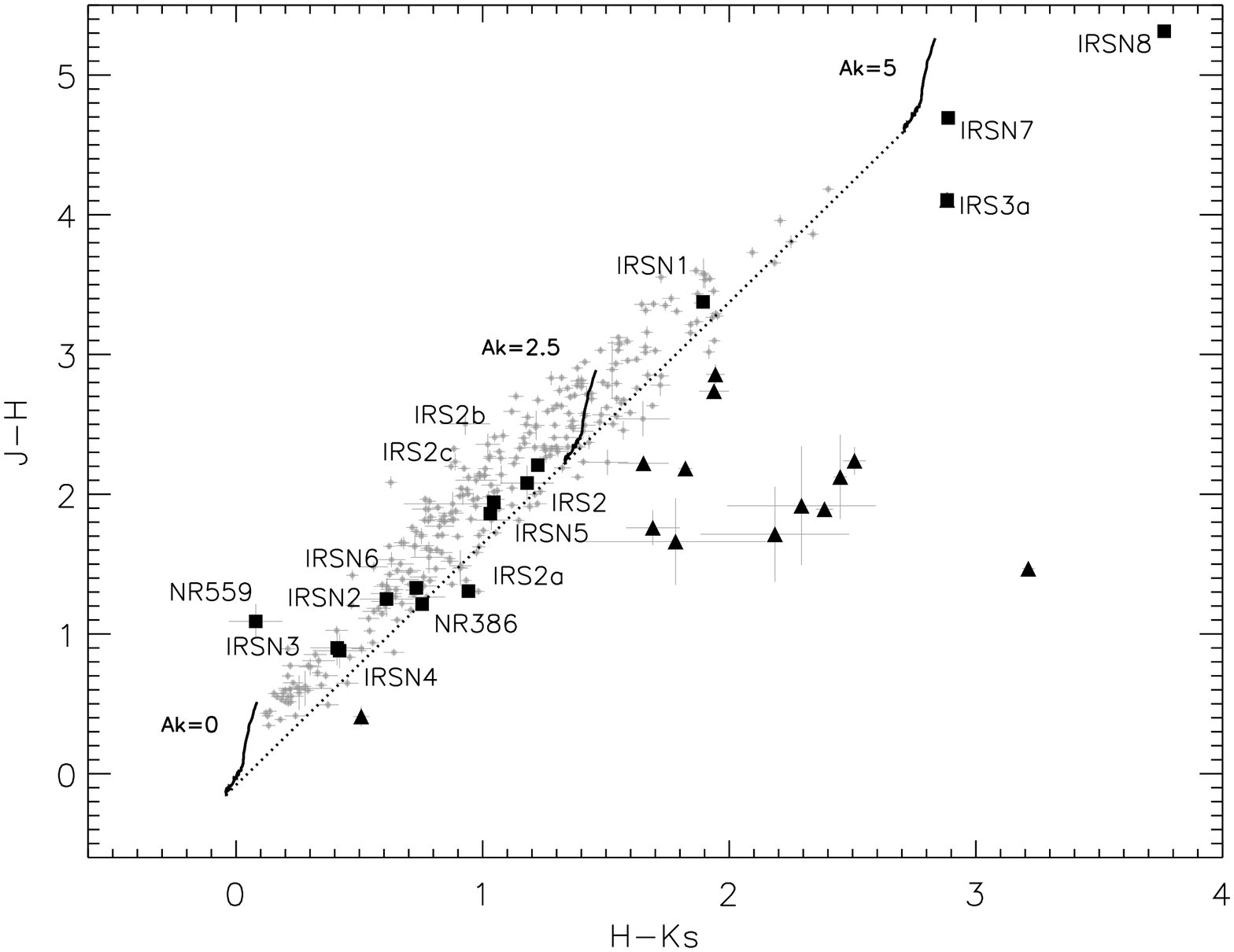}
\caption{{\emph Left:} $H-K_{\rm{s}}$ vs $K_{\rm{s}}$ color-magnitude diagram of W3 Main.  Plotted as squares are the candidate massive stars identified by \citet{Ojha04}, in this work and the two stars identified by \citet{Navarete11}.  Overplotted as a dashed line is the main sequence isochrone with an age of 1 Myr from \citet{LeJeune01} with the location of the spectral types between O3V and B3V are indicated. 
The two diagonal dotted lines show the reddening lines for an O3V and a B3V star using the extinction law of \citetalias{Indebetouw05}.  Additionally, the reddening vectors for $A_{K_{\rm{s}}}$=1 mag from \citetalias[(steep vector)]{Indebetouw05} and \citetalias[(shallow vector)]{Nishiyama09} are plotted. {\emph Right: $J-H$, $H-K_{\rm{s}}$ color-color diagram of W3 Main. The plotting symbols and lines are the same as in the CMD. The stars with an infrared excess, located more than 3$\sigma$ away from the reddened main sequence are plotted as triangles.}  \label{fig:cmd}}
\end{figure*}

\subsubsection{Multi-object spectroscopy}

The MOS spectra were reduced with a modified version of  \emph{lucired}, which is a collection of IRAF routines developed for the reduction of LUCI MOS spectra.  First the raw frames were corrected for the tilt of the slit using a spectroscopic sieve mask as well as for distortion through an imaging pinhole mask. After correction, the slits were oriented horizontally on the frame and were extracted to be treated separately. The science data were corrected by a normalized flat field. The flat field was normalized in the spatial direction for every slit to remove illumination effects as well as the slit response. 

The long slit spectra were reduced using standard IRAF routines and no distortion or tilt correction was applied. The flat field was normalized by dividing with the mean of the flat field as the spatial variation in the counts of the flat field was very low.

For both observing modes, the wavelength calibration was performed using the Ar and Ne arc wavelength calibration frames. After the wavelength calibration,  the sky background was subtracted using the  procedure by \citet{Davies07} to remove the OH line residuals. Finally, the 1D spectra were extracted using the iraf task \emph{doslit} and combined. 

The removal of the telluric absorption lines was performed in two steps. First the telluric absorption lines in the standard star spectrum were removed using a high-S/N telluric reference spectrum\footnote{obtained at NSO/Kitt Peak Observatory},  providing a first rudimentary correction in order to reduce the absorption contamination of the telluric standard star. The \brg\ absorption line  was  fitted with a Lorentzian profile. The best fitting profile of \brg\ was removed from the original standard star spectrum, resulting in a atmospheric transmission spectrum taken at the same airmass as the science observations. Finally, the IRAF task \emph{telluric} was used to correct  the science observations for the telluric absorption lines.  This procedure was applied to the standard star observations before and after the science observations as well as in the two different slits.  The correction showing the least telluric residuals was chosen as the final reduction. Finally, the spectra are normalized by fitting a spline function to the continuum.

\section{Results}

\subsection{Photometry}

The final reduced 3-color $JHK_{\rm{s}}$ image is displayed in Fig. \ref{fig:W3JHK}.  Apart from the stellar content, also several patches of extended emission can be seen. These areas coincide with the location of the \HII\ regions in the radio maps \citep{Tieftrunk97}. 
W3 A is the brightest \HII\ region in the near-infrared in the north-east of the image. The two diffuse \HII\ regions W3 K and W3 J can be identified in scattered light in the $J$-band due to the low extinction. Several dark dust lanes can be seen as well, showing that the extinction is highly variable.

The $K_{\rm{s}}$, $H-K_{\rm{s}}$ color-magnitude diagram  (CMD) and $J-H$, $H-K_{\rm{s}}$ color-color diagram (CCD) confirm this large extinction variation over the observed field (Fig. \ref{fig:cmd}).  The CMD and CCD  show all sources with a detection in all three bands and  brighter than $K_{\rm{s}}$ = 15 mag.  In the CMD, no clear reddened, cluster main sequence can be detected.  Apart from a foreground main sequence (at $H-K_{\rm{s}}$ $\approx$ 0.2 mag), an enormous spread in $H-K_{\rm{s}}$ color of the sources in W3 Main is observed, caused by large extinction variations inside W3 Main. The CCD of the bright stars with $K_{\rm{s}}$ $\le$ 15 mag confirms these huge extinction variations, as the location of the reddened main sequence stars (solid lines) is populated to at least $A_{\rm{K_s}}$ = 5 mag. Several sources are located to the right of the reddened main sequence,  due to  an infrared excess, most likely originating in a circumstellar disk (see Fig. \ref{fig:cmd}). Overplotted in the CMD is the location of the upper end of the un-reddened main sequence 1 Myr isochrone \citep{LeJeune01}, with the spectral types annotated as well as the reddening vectors from the \citet[hereafter Ind05]{Indebetouw05} and \citet[hereafter NI09]{Nishiyama09} extinction laws. The candidate massive stars in W3 Main are expected to be located in between the reddening lines departing from a B3V and  a O3V star respectively (diagonal lines in the CMD).

\begin{deluxetable*}{llllrrrrrrr}
\tablewidth{0pt}
\tablecaption{Catalogue of the massive stars in W3 Main \label{tab:stars}}
\tablehead{
\colhead{Object\tablenotemark{a}} &
\colhead{\HII\ } &
\colhead{$\alpha$ (J2000)} &
\colhead{$\delta$ (J2000)} &
\colhead{$J$} &
\colhead{$H$} &
\colhead{$K_{\rm{s}}$} \\
\colhead{} &
\colhead{region} &
\colhead{(h m s)} & 
\colhead{($^\circ$\ \arcmin\  \arcsec)} &
\colhead{mag} &
\colhead{mag} &
\colhead{mag} 
}
\startdata
IRS2      				& W3 A	& 02:25:44.3	& +62:06:11.4	& 12.23 $\pm$ 0.10	&	10.15 $\pm$ 0.07	&	8.97   $\pm$ 0.08		 \\
IRS2a      				& W3 A	& 02:25:43.3	& +62:06:15.2	& 12.90 $\pm$ 0.03	&	11.60 $\pm$ 0.01	&	10.65 $\pm$ 0.02		\\
IRS2b       			& W3 A	&02:25:41.7	& +62:06:24.2	& 13.92 $\pm$ 0.03	&	11.71 $\pm$ 0.01	&	10.49 $\pm$ 0.02  		\\
IRS2c      				& W3 A	& 02:25:47.1	& +62:06:13.0	& 13.58 $\pm$ 0.03	&	11.64 $\pm$ 0.01	&	10.59 $\pm$ 0.02  			\\
IRS 3a       			& W3 B	& 02:25:37.8	& +62:05:51.8	& 19.55 $\pm$	0.03 &	15.44 $\pm$ 0.01	&	12.56 $\pm$ 0.02  		\\
IRS4					& W3 C	& 02:25:31.0	& +62:06:20.6	&		---		&	16.45 $\pm$ 0.02	&	13.41 $\pm$ 0.02		\\ 
IRS5					& W3 M   & 02:25:40.8	& +62:05:52.3	&		---		&			---		&	13.16 $\pm$ 0.02	\\
IRS7					& W3 F	& 02:25:40.5	& +62:05:39.8	&		---		&	17.07 $\pm$ 0.02	&	13.90 $\pm$ 0.02		\\ 
IRS N1     				& W3 E	& 02:25:35.1  	& +62:05:34.5	& 17.28 $\pm$	0.02 &	13.91 $\pm$ 0.01	&	12.01 $\pm$ 0.02		\\ 
IRS N2				& W3 H	&02:25:32.6 	& +62:06:59.6	& 11.74 $\pm$	 0.10	&	10.49 $\pm$ 0.07	&	9.88 $\pm$ 0.08		\\ 
IRS N3      			& W3 J	& 02:25:27.4	& +62:03:43.2	& 11.25 $\pm$ 0.10	&	10.35 $\pm$ 0.07	&	9.94 $\pm$ 0.08		\\ 
IRS N4     	 			& W3 K	& 02:25:44.8	& +62:03:41.0	& 10.31 $\pm$	0.10	&	9.43   $\pm$ 0.07	&	9.01   $\pm$	0.08		\\
IRS N5\tablenotemark{b}	& 		& 02:25:32.7	& +62:05:08.1	& 14.61 $\pm$	0.03	&	12.75 $\pm$ 0.01	&	11.72 $\pm$ 0.02			\\
IRS N6\tablenotemark{b} & W3 A?	& 02:25:47.4	& +62:06:55.3	& 15.04 $\pm$	0.02 &	13.71 $\pm$ 0.02	&	12.97 $\pm$ 0.02		\\
IRS N7\tablenotemark{b}	&      	& 02:25:40.6	& +62:05:46.8	& 17.99 $\pm$	0.02	&	13.30 $\pm$ 0.01	&	10.41 $\pm$ 0.02		\\
IRS N8\tablenotemark{b}	&		& 02:25:26.3	& +62:06:07.4	& 20.99 $\pm$	0.30	&	15.67 $\pm$ 0.01	&	11.91 $\pm$ 0.02		\\
NR559\tablenotemark{c}	& 		& 02:25:28.1	& +62:05:39.6  & 9.62   $\pm$  0.10 &       8.53  $\pm$ 0.07	&        8.45 $\pm$ 0.08   	\\               	
NR386\tablenotemark{c}	&		& 02:25:37.5	& +62:05:24.5	& 12.36 $\pm$  0.03	 &      11.15 $\pm$  0.01	&	10.39  $\pm$ 0.02	
\enddata
\tablenotetext{a}{taken from \citet{Ojha04}}
\tablenotetext{b}{naming convention discussed in main text}
\tablenotetext{c}{Taken from \citet{Navarete11}}
\end{deluxetable*}

\citet{Ojha04} identified 13 candidate OB stars which are bright near-infrared point sources without infrared excess and associated with \HII\ regions detected in the radio (IRS2 -- IRS N4 in Table \ref{tab:stars}).  The two candidate massive stars (IRS4 and IRS7)  have only $H$- and $K_{\rm{s}}$ detections, but are added to the CMD, and are not plotted in the CCD. IRS5 is only detected in $K_{\rm{s}}$.  To identify massive Young Stellar Objects (YSOs) we added bright sources possessing an infrared excess and added two more bright sources in the field to complete the target list for the spectroscopic observations (IRS N5 - IRS N8 in Table. \ref{tab:stars}).  Recently, \citet{Navarete11} published $K$-band spectra of 4 sources in W3 Main. Two of them (NR386 and NR559)  are added to our sample and included in our analysis. The two other stars, IRS N3 and IRS N4, in common with our sample, are used as a cross check of the spectral type determination.

A comparison between our LUCI photometry and those of \citet{Ojha04} and 2MASS shows significant differences. The LUCI photometry of the blue and isolated sources is fully consistent with the photometry by \citet{Ojha04} and 2MASS.  However, for the red sources  (IRS4 and IRS7), as well as sources contaminated by strong background emission (IRS2a and IRS2b), a significant difference (in $H$ up to 0.7 mag and in $K_{\rm{s}}$ up to 0.9 mag) is found.  All these sources are located inside  \HII\ regions and the lower resolution images of \citet{Ojha04} as well as 2MASS possibly include more nebular contribution, altering the magnitudes. Additionally, for the extremely red sources, a possible small color term in the  conversion between the different photometric systems, for either our or the photometry of \citet{Ojha04},  could also contribute to the observed differences.  Due to the finer pixel scale and better seeing of our LUCI observations we consider our LUCI photometry more reliable. 

The CMD (Fig. \ref{fig:cmd}) can be used to estimate the spectral type of the candidate massive stars by comparison with  the absolute magnitudes predicted by the 1 Myr isochrone of  \citet{LeJeune01} and adopting a distance of 1.95 kpc  \citep{XuW306}. The 1 Myr isochrone has been chosen as it likely represents the age of the region better than the ZAMS (see also Sect. 4 for a discussion on the age).  Using the relation between T$_{\rm{eff}}$  and spectral type from  \citet{Martins05} for the O stars and \citet{Kenyon95} for the B stars we derived the matching spectral types.

\begin{figure*}
\plotone{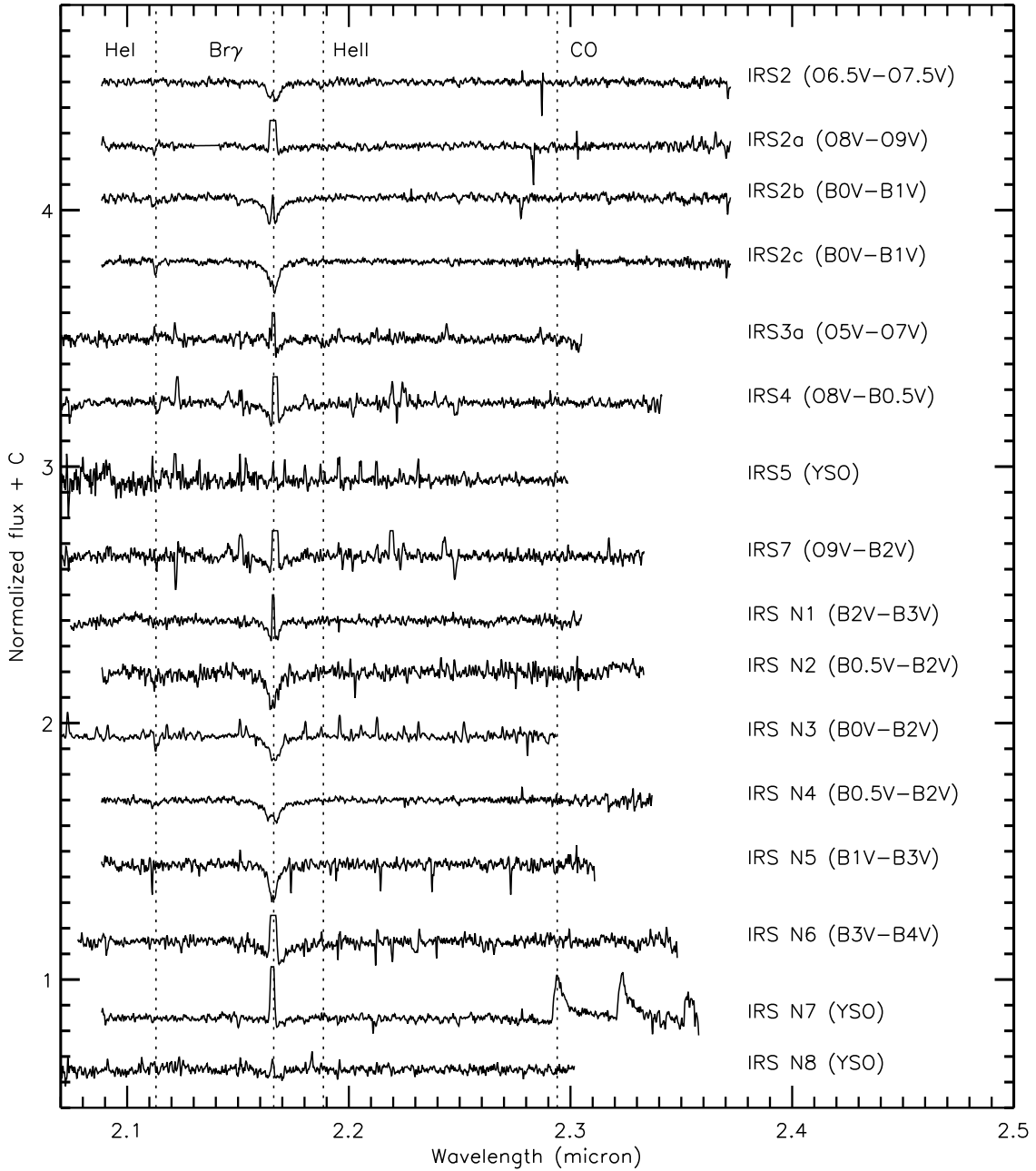}
\caption{Normalized $K$-band spectra of the massive stars in W3 Main as taken with the multi-object-mode of LUCI. Annotated with dashed lines are the spectral features which are important for classification of the stellar spectra. For the details of the classification see the main text. \label{fig:spectra}} 
\end{figure*}

The so derived photometric spectral types are very sensitive to the chosen extinction law. Especially for the high-extinction sources, a different  extinction law changes  the spectral type significantly by up to 3-4 subtypes. We have tested several extinction laws and compared them to the observed CCD (Fig. \ref{fig:cmd}). The extinction laws of \citet{Rieke85}, \citetalias{Indebetouw05} and \citetalias{Nishiyama09} provide a reasonable fit to the observed CCD, having  a very similar $E(J-H)/E(H-K_{\rm{s}})$ value of $\sim$ 1.73 (1.69, 1.73 and 1.76, respectively). The  extinction laws of \citet{Cardelli89}, \citet{Fitzpatrick99} and \citet{Roman07} do not match the observed distribution of the stars in the CCD, showing a too shallow, $E(J-H)/E(H-K_{\rm{s}})$ = 1.21 \citep{Cardelli89},  and , $E(J-H)/E(H-K_{\rm{s}})$ = 1.52 \citep{Roman07},  or too steep slope , $E(J-H)/E(H-K_{\rm{s}})$ = 1.84 \citep{Fitzpatrick99} of the reddened main sequence.  In the CMD and CCD  (Fig. \ref{fig:cmd}) we have plotted the extinction law of \citetalias{Indebetouw05}.

 The other parameter of the extinction law,  the total over selective extinction ($R_{\lambda} = A_{\lambda}/E(H-K_{\rm{s}}$)), is important when absolute values of $A_{\rm{K_s}}$ are derived e.g. to place the objects in the HRD or derive a photometric spectral type.  $R_{\rm{K_s}}$ is  difficult to determine from $JHK_{\rm{s}}$ data only. \citetalias{Indebetouw05}, \citet{Rieke85} and \citetalias{Nishiyama09} provide different values for $R_{\lambda}$, while having a similar slope of the near-infrared extinction law. 
 \citetalias{Indebetouw05} derive $R_{\rm{K_s}}= A_{\rm{K_s}}/E(H-K_{\rm{s}}) = 1.82 \pm 0.09$, \citet{Rieke85} find $R_{\rm{K_s}}=  1.78$, while \citetalias{Nishiyama09} derive $R_{\rm{K_s}}  = 1.44 \pm 0.01$.  
 This parameter can only be constrained when the absolute magnitude of a reddened source is known.   The effect of this parameter on the HRD will be discussed in Sect. 4.1. 

Table \ref{tab:sptypes} shows the resulting photometric spectral types for the extinction laws of \citetalias{Indebetouw05} and \citetalias{Nishiyama09}. For the low extinction sources, $A_{\rm{K_s}} <$ 1.5 mag, both laws result in  similar photometric spectral types. For the highly embedded sources such as IRS4 and IRS7, the two spectral types differ dramatically, up to 4 subtypes.

Apart from the candidate OB stars inside the \HII\ regions,  other candidate OB stars can be found in the CMD. \citet{Navarete11} identified one of the blue sources  as a O7V star, not related to any \HII\ region. This would suggest that all the other bright stars could well be O stars, while the stars around $K_{\rm{s}}$=13 mag and $H-K_{\rm{s}} \approx$ 2 mag are most likely candidate B stars and more widely distributed across the field of W3 Main. 

The CCD reveals a number of  infrared excess sources. Taking into account the photometric errors, we found 15 sources located more than 3$\sigma$ from the reddening line. The majority of the sources are faint ($K_{\rm{s}} \approx$ 14 mag) and are most likely intermediate mass PMS stars showing circumstellar disk emission.  Among the brighter sources, two are spectroscopic targets: IRS3a and IRS N8, with additionally IRS2a and IRS7 located very close to the 3$\sigma$ line. The $K$-band spectra of  these sources are thus crucial to establish the true nature of their IR excess.

\begin{deluxetable*}{lrrrrr}
\tablewidth{0pt}
\tablecaption{Spectral types of the massive stars in W3 Main \label{tab:sptypes}}
\tablehead{
\colhead{Object\tablenotemark{a}} &
\colhead{S/N ratio}&
\colhead{Photom.} &
\colhead{Spectro.}&
\colhead{$A_{\rm{K_s}}$\tablenotemark{b}}\\
\colhead{} &
\colhead{} &
\colhead{Class.$\tablenotemark{a}$} &
\colhead{Class.} &
\colhead{mag} 
}
\startdata
IRS2      	& 135 	&O4V/O4.5V	&	O6.5V -- O7.5V		& 2.32 $\pm$ 0.39		 \\
IRS2a      	& 125	&O9V/B0V	&	O8V -- O9V	& 1.90 $\pm$ 0.34	 	\\
IRS2b       & 130	&O7V/O9V	&	B0V -- B1V		& 2.41 $\pm$ 0.40		\\
IRS2c      	& 155	&O8V/B0V	&	B0V -- B1V		&  2.08 $\pm$  0.36		\\
IRS 3a      & 75		&O5V/O8V	&	O5V	-- O7V	& 5.42 $\pm$  0.79	\\
IRS4		& 100 	&O6V/B0V	&	O8V -- B0.5V		& 5.69 $\pm$ 0.83	\\ 
IRS5		& 50 	&---	&	YSO			&		---		\\
IRS7		& 60		&O7V/B0V	&	O9V -- B2V		& 5.93 $\pm$ 0.86	\\ 
IRS N1     & 105	&O8V/B0V	&	B2V -- B3V		& 3.60 $\pm$ 0.53	\\ 
IRS N2	& 60		&O9V/O9V	&	B0.5V -- B2V 			&1.28 $\pm$ 0.27	\\ 
IRS N3     & 110	&B0V/B0V	&	B0V  -- B2V			& 0.92 $\pm$ 0.14	\\ 
IRS N4     	& 140	&O7V/O7V	&	B0.5V -- B2V	& 0.94 $\pm $ 0.24		\\
IRS N5	& 80  	&B1V/B1V	&	B1V -- B3V		& 2.04 $\pm$ 0.35		\\
IRS N6	& 70 	&B4V/B4V	& 	B3V -- B4V		& 1.47 $\pm$ 0.22	\\
IRS N7	& 110	&$>$O3V	&	YSO		& 	 ---	\\
IRS N8	& 70	&$>$O3V/O3V  	&	YSO		& 	--- 	\\
NR559	&  ---    	& O7V/O7V	& 	O7V		& 0.33  $\pm$   0.20	\\               	
NR386	&  ---		&O9V/B0V	&	B0V -- B2V & 1.54 $\pm$    0.23
\enddata
\tablenotetext{a}{Spectral types derived with the \citetalias{Indebetouw05} and \citetalias{Nishiyama09}  extinction laws}
\tablenotetext{b}{Values derived using the \citetalias{Indebetouw05} extinction law, when adopting the \citetalias{Nishiyama09} law, the values need to be increased by a factor 0.79$\times E(H-K_{\rm{s}})$}
\end{deluxetable*}

\subsection{Spectral classification}

$K$-band spectroscopy of the candidate massive stars provides  more reliable spectral types, based on the presence of photospheric absorption and emission lines, than  photometry alone. Additionally, it reveals the nature of the infrared excess sources detected in the CCD.
 
 The $K$-band spectra of the massive stars are shown in Fig. \ref{fig:spectra}. The majority of the spectra shows \brg\ absorption with a narrow emission component. The absorption has a  photospheric origin, while the narrow emission line  is emitted by the surrounding \HII\ region.  Due to variations of the nebular emission on small spatial scales, we made no attempt to correct the spectra for nebular emission. Other lines present in OB star atmospheres are \ion{He}{1} 2.113 \micron,  the \ion{N}{3} complex at 2.115 \micron\ (IRS2a and IRS3a) and \ion{He}{2} at 2.1185 \micron\ (IRS2). 
 
The LUCI spectra  are compared with high-resolution reference spectra of optically classified O and early B stars from \citet{Hanson05} and \citet{Ostarspec05}. Reference spectra  of mid- and late-B stars were added from  \citet{Hanson96}. The reference spectra were degraded to the resolution of the LUCI spectra and artificial noise was added to make the  spectra comparable in S/N. A visual comparison is performed to determine the best matching spectral type.

As an additional check, we determined the spectral types by comparing the measured equivalent widths (EW) of the \brg, \ion{He}{1} and \ion{He}{2} absorption lines with the observations of  \citet{Hanson96} and theoretical predictions of \citet{Lenorzer04Model}. Since the \brg\ absorption lines are contaminated by the nebular emission, special care has been taken to measure the EW of \brg\ reliably.  We fit a Moffat profile to the absorption wings  of \brg\ which are unaffected by the narrow nebular emission line, providing a more reliable EW determination.  The Moffat fit resembles the observed line profile well and reproduces the line shapes of IRS N3 and IRS N5, which are unaffected by nebular emission.  The two methods deliver very similar spectral types.
\citet{Navarete11} obtained $K$-band spectra of IRS N3 and IRS N4, and derived the same spectral types as we derive based on our LUCI spectra.

\begin{figure*}
\plottwo{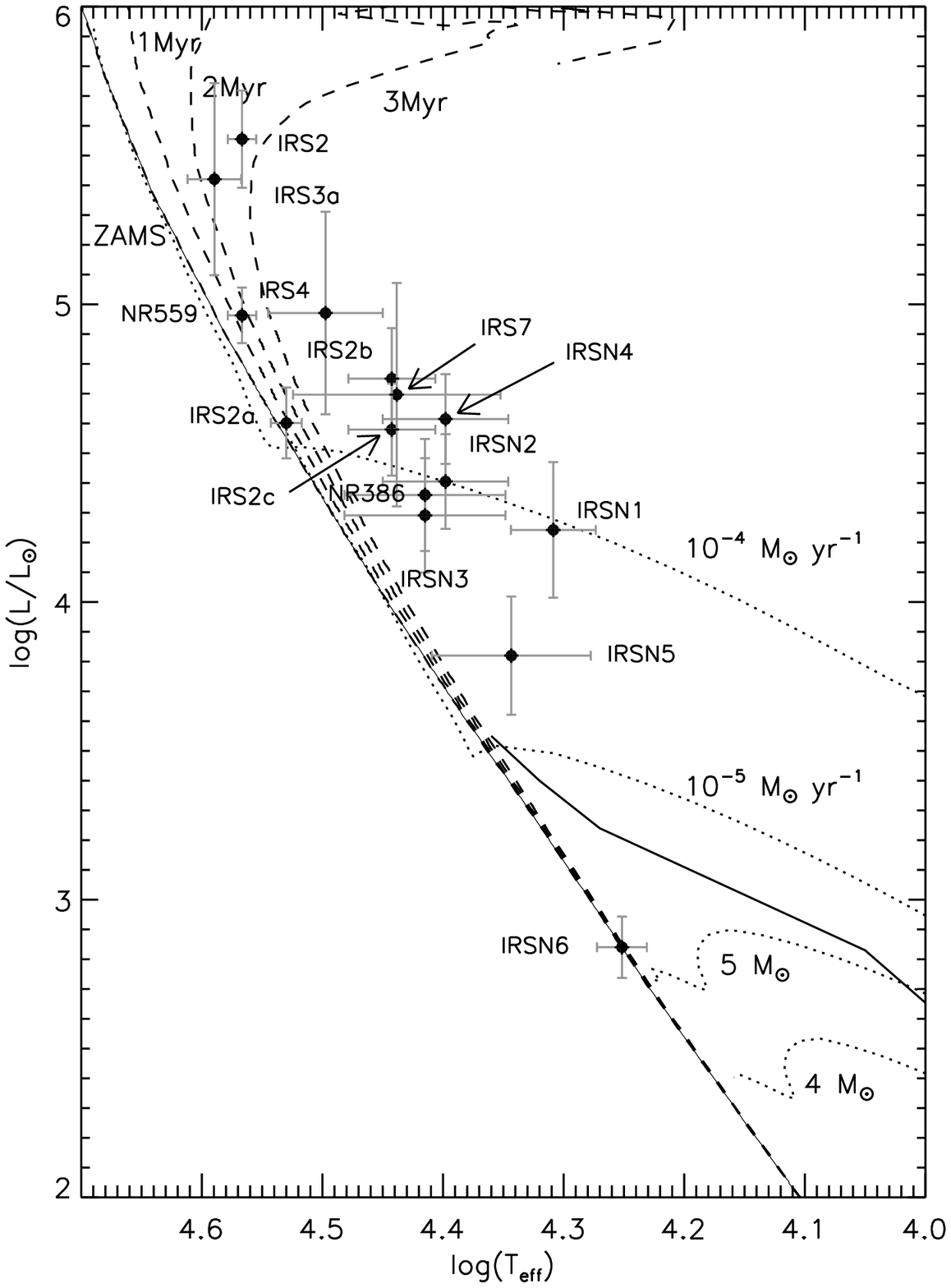}{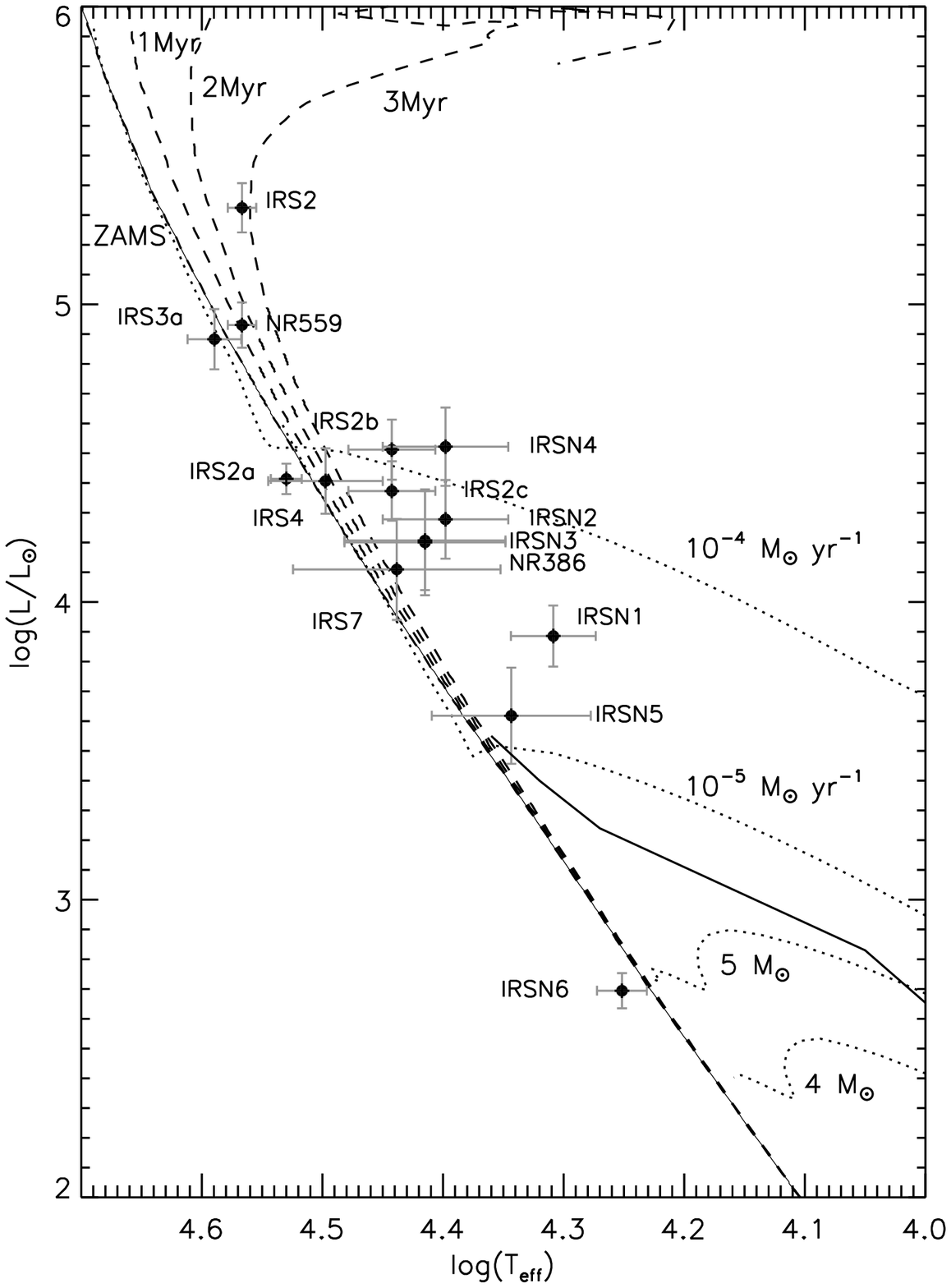}
\caption{HRD  of the massive stars  in W3 Main. The dashed lines represent the main-sequence isochrones from \citet{LeJeune01} for 1, 2 and 3 Myr. The dotted lines show the PMS evolutionary tracks for a 4 and 5 $M_{\sun}$ star \citep{Siess00} as well as high-mass protostellar evolution tracks for accretion rates of  10$^{-4}$ and 10$^{-5}$ $M_{\sun}$ yr$^{-1}$ from \citet{Hosokawa09}.  The solid line represents the theoretical birth-line, above which a  PMS star is not detectable in the optical \citep{Palla90}. 
\emph{left}: The stars are de-reddened using the extinction law of \citetalias{Indebetouw05} extinction law.  
\emph{right:} The same  HRD, but  using the extinction correction of \citetalias{Nishiyama09}. See also discussion in the text. \label{fig:hrd}}
\end{figure*}

As shown in Fig. \ref{fig:cmd}, four stars with spectra (IRS2a, IRS3a, IRS N7 and IRS N8)  show evidence for a near-infrared excess in the CCD. Two of them, IRS N7 and IRS N8, display an emission line spectrum. Additionally, the $K$-band spectrum of the deeply embedded source IRS5 also has these characteristics.  These spectra show \brg\ emission and  IRS N7 also shows the CO bandhead emission at 2.3 \micron, most likely originating from dense, hot circumstellar material around the star \citep{Coletter04,Blum04}. These emission line spectra, as well as rising continua, suggest that these objects are massive YSOs,  object surrounded by remnant accretion disk, possibly as a result of their formation \citep{Brgspec06}.

The  other sources with a near-infrared excess, IRS2a and IRS3a, however, do not show a spectrum indicative of circumstellar matter.  Their $K$-band spectra show photospheric absorption and emission lines and they are classified as  O8V-O9V (IRS2a) and O5V-O7V (IRS3a). Possibly, free-free emission of the bright \HII\ region could be the cause of their IR excess.  This shows the importance of spectroscopy to identify the stellar content reliably and disentangle the massive YSO candidates from the sources contaminated by nebular emission. 

Veiling by free-free emission makes the absorption lines more difficult to detect and could increase the uncertainty of  the spectral type determination. In both IRS2a and IRS3a, the \ion{N}{3} complex at 2.115 \micron\ is detected, narrowing down the spectral type determination to O5V-O9V \citep{Hanson96} and making the effect of veiling on the spectral type determination less important.

Using LUCI MOS spectroscopy, we have determined the spectral type of 13 OB stars in W3 Main as well as identified 3 massive YSOs. Adding the two sources from \citet{Navarete11} to our sample results in a total of 15 OB stars. The final derived spectral types are given in Table \ref{tab:sptypes} and vary from O5V - O7V for IRS3a to B3V-B4V for IRS N6.  

When we compare the spectral types with those derived from the photometry (Sect. 3.1), it turns out that the photometry usually predicts earlier spectral types. For some objects, like IRS2a, IRS3a, the spectral types derived with both methods agree well. However, in many cases the spectral types are rather different, e.g. for IRS2, IRS N1 and IRS N4.  In the case of IRS N4, the photometric spectral type is O7V (for both extinction laws), for which the  \ion{N}{3} emission line at 2.115 \micron\   and the \ion{He}{2} absorption line at 2.185 \micron\  are expected. However, in the  spectrum of IRS N4 these lines are not observed, resulting in a much later spectral type (B0.5V - B2V).

\subsection{HR-diagram}

After spectroscopic classification, the objects can be placed in the HRD. The intrinsic $H-K_{\rm{s}}$ colors as well as the bolometric corrections are taken from  \citet{Martins06} for the O stars and \citet{Kenyon95} for the B stars. First, the $E(H-K_{\rm{s}})$ is calculated and converted into $A_{\rm{K_s}}$.  For the determination of  $A_{\rm{K_s}}$, the two extinction laws \citepalias{Indebetouw05,Nishiyama09} provide  different values. In table \ref{tab:sptypes}, the values of $A_{\rm{K_s}}$ are given for  the \citetalias{Indebetouw05} extinction law. To obtain the values using the \citetalias{Nishiyama09} extinction law, a factor of 0.79$\times E(H-K_{\rm{s}})$ needs to be applied.

After accounting for  extinction, and applying the distance modulus and the bolometric correction, the luminosities of the massive stars in W3 Main are derived and they are placed in the HRD (Fig \ref{fig:hrd}). The left panel shows the HRD with the extinction correction by \citetalias{Indebetouw05}  ($R_{\rm{K_s}}$  = 1.82), while the  right diagram shows the correction  by \citetalias{Nishiyama09} ($R_{\rm{K_s}}$ = 1.44).  The error budget of the luminosity is dominated by the errors on the extinction coefficients, while the error in T$_{\rm{eff}}$ is given by the uncertainty of the spectral classification, which in turn results in an extra uncertainty in the bolometric correction.

Overplotted in the HRDs are the zero age main sequence (ZAMS) and  main sequence isochrones from 1 to 3 Myr from \citet{LeJeune01}, the PMS isochrones for intermediate mass stars  \citep[$M < 5 M_{\sun}$,][]{Siess00} as well as the theoretical birth-line \citep{Palla90}. Overplotted also are the high-mass protostellar evolutionary tracks from \citet{Hosokawa09} for 10$^{-5}$ and 10$^{-4} M_{\sun}$ yr$^{-1}$ accretion rates.

Comparison between the two HRDs shows that, on average, the stars are closer to the main sequence  in the right diagram with the \citetalias{Nishiyama09} extinction correction.
In the left diagram, with the \citetalias{Indebetouw05}  law, most of the sources are located to the right of the main sequence. Only IRS2a and IRS N6 fit better  the main sequence in this diagram, while still in good agreement with the main sequence location in the \citetalias{Nishiyama09} case.  The very deeply embedded sources IRS4 and IRS7 are located  closer to the main sequence in the diagram with the  \citetalias{Nishiyama09} law.  Several sources, like IRS N1, IRS N4 and IRS2 remain to the right of the main sequence in both diagrams. 

To clarify the description of the HRD, we divide the HRD in three areas: the upper HRD with log $L/L_{\sun}$ $>$ 5,  where IRS2 and IRS3a are located. The middle area of the  HRD (3.5 $<$ log $L/L_{\sun}$ $<$ 5), is where most of the stars are located. The lower HRD (log $L/L_{\sun}$ $<$ 3.5) is the area of the intermediate mass (PMS) stars where  IRS N6 is located close to the PMS turn-on region.

Two stars are located in the upper  HRD, IRS2 and IRS3a, where the theoretical isochrones indicate that the most massive stars already show significant evolution  after a few Myr. As already noted in Sect. 3.1,  IRS2 has a significant difference between the photometric and spectroscopic spectral type. It is the most luminous object in W3 Main, but not the earliest type star (O6.5V - O7.5V). This is illustrated by its location to the right of the ZAMS,  more consistent with the 2 - 3 Myr isochrones. The foreground extinction is not very extreme, therefore the  location of IRS2 in the HRD does not vary a lot for different  extinction laws.

IRS3a (O5V - O7V), however, is very reddened (A$_{\rm{K_s}}$=5.42 $\pm$ 0.79 mag) and its location in the HRD depends  strongly on the adopted extinction law. Additionally, the presence of a near infrared excess (Fig. \ref{fig:cmd}) makes its luminosity more uncertain. However, the effective temperature determination is more reliable and independent of extinction and excess. This allows us to place an upper limit on the age of 3 Myr, as stars on older isochrones will be of later spectral type than IRS3a.

In the middle HRD, the position of the sources most consistent - within 3$\sigma$ - with their expected main sequence location. For spectral types between O9V and B1V, where most of the objects are found, the change in T$_{\rm{eff}}$ and $K$-band magnitude is large between e.g. O9V and B0V. However, the the change in observable spectral features is not that dramatic.  This is reflected in the large error bars of e.g. IRS N5 and IRS7.   IRS N1 and  IRS N4, however, are the furthest away from the main sequence. As their extinction is relatively low, their position does not change much adopting different extinction laws. Below we discuss possible scenarios explaining this offset.

A mis-classification of the spectral type is unlikely. Similarly as described in Sect. 3.2, the observed spectra of IRS N1 and IRS N4 are not compatible with the expected spectral type derived from their luminosity  (O7V and  B1V). 

Another explanation for the offset in the HRD is that the sources are at different distances. Calculating the spectrophotometric distance of the stars, assuming that they all are on the main sequence, would result in a distance between 1 and 3.4 kpc suggesting that subregions of  W3 Main would be 1 kpc in the fore- or background. \citet{Heyer98} show that the entire W3/W4/W5 molecular cloud complex forms a connecting structure, both in spacial projection and velocity, suggesting that the entire complex is at the same distance \citep[see also][]{Megeath08}. Therefore, we consider it highly unlikely that the offsets in the HRD can be explained by different distances for the individual sources.

Other  effects, however, can be responsible for this offset in the HRD. 
Most of the massive stars are observed to be binary stars \citep[e.g.][]{Bosch01,Apai07}. An equal mass binary  would result in a mismatch in the observed luminosity by a factor of 2 ($\Delta$ log $L/L_{\sun}$  = 0.3 dex),  with a multiple star system increasing the offset even more. Both sources would have to be a system of 5-6 equal mass stars to explain their location above the main sequence, though.

Stellar evolution can also be the reason why these objects are located away from the main sequence. The location of IRS N4 would be compatible with a 10 Myr main sequence track, while IRS N1 would require an even older isochrone. This is highly unlikely, in particular for IRS N1, as the star is located inside an UC\HII\ region, which is still very young \citep[$\sim10^5$ yrs,][]{WoodIRAS89}. IRS N4 is the central star of a diffuse and more evolved \HII\ region; as shown by \citet{Lada03} a timespan of 10 Myr would be enough to disperse the surrounding gas.

Another possible explanation for the HRD loci of the stars is that they are still in the PMS phase. 
Above the ``birth-line" no optically visible PMS stars are expected \citep{Palla90}.  However, using deep near-infrared observations we might start to probe the PMS phase of  massive stars. Theoretical modeling of  the protostellar evolution predicts that high-mass  protostars might have a similar evolution to intermediate PMS stars towards the main sequence \citep{Hosokawa09}.

 The exact path of the protostars in the HRD depends on the accretion rate. Overplotted in Fig. \ref{fig:hrd} are two  tracks for accretion rates of  10$^{-4} M_{\sun}$ yr$^{-1}$ and 10$^{-5} M_{\sun}$ yr$^{-1}$. The location of IRS N1 would be consistent with  an accretion rate between 10$^{-5}$  and 10$^{-4} M_{\sun}$ yr$^{-1}$, while IRS N4 would require a slightly higher accretion rate.

The location of IRS N1 and IRS N4 are consistent with the very last stages of the PMS evolution where the stars are contracting to their main-sequence temperature and luminosity. In this contraction phase, the accretion luminosity only contributes 20\% in the case of the 10$^{-4} M_{\sun}$ yr$^{-1}$ and 10\% 10$^{-5} M_{\sun}$ yr$^{-1}$ track  respectively. The absorption lines in the K-band spectra of IRS N1 and IRS N4 as well as the lack of near-infrared excess are consistent with the stellar contribution being the dominant source of luminosity.

Surprising for those young objects however, is the lack of infrared emission coming from a circumstellar disk. The presence of a disk would make these objects similar to e.g. IRS N7 or IRS N8, which show an emission line spectrum and a near-infrared excess. An explanation of the lack of disk emission could be the fast dispersal of the disk by FUV photons of the central star or external FUV photons from other cluster members in W3 Main (see Section 4.1).

Summarizing, the spectral classification of the massive stars in W3 Main allows the comparison with stellar evolution models in the HRD. The location of the most massive stars in the upper HRD suggests that IRS2 has an age of 2-3 Myr and IRS3a is slightly younger, while a speculative explanation of the displacement of  IRS N1 and IRS N4 in the HRD could be high-mass PMS evolution. However,  due to the large impact a change of extinction law has on the appearance of the HRD, it is  difficult to assess with confidence that these stars are in the PMS phase. To prove this hypothesis, a good determination of the stellar surface gravity, by means of modeling the absorption lines of  high signal-to-noise spectra is needed to show that they have  larger radii than main sequence stars.

\begin{deluxetable*}{lcllrr}
\tablewidth{0pt}
\tablecaption{Ionizing content of the \HII\ regions in W3 Main\label{tab:hii}}
\tablehead{
\colhead{\HII\ } &
\colhead{log $N_{\rm{Ly}}$ } &
\colhead{Radio} &
\colhead{Stars} &
\colhead{NIR} &
\colhead{log $Q_{\rm{0}}$}\\
\colhead{region } &
\colhead{s$^{-1}$} &
\colhead{Sp. type } &
\colhead{} &
\colhead{Sp. type} &
\colhead{s$^{-1}$}
}
\startdata
W3 A		& 48.93	& O6V	& IRS2	& O6.5V -- O7.5	 & 48.44 -- 48.80 \\
W3 B		& 48.43	& O7.5V	& IRS3a	& O5V -- O7V	& 48.63 -- 49.26	 \\
W3 C		& 47.46	& O9.5V	& IRS4	& O8V -- B0.5V	& 47.00 -- 48.29	\\
W3 E		& 46.50	& B1V	& IRS N1	& B2V -- B3V	&  $\leq$ 46.10	\\
W3 F		& 47.04	& B0V	& IRS7	& O9V -- B2V	& $\leq$ 47.56	\\
W3 H		& 47.85	& O9V	& IRS N2	& B0V -- B2V	& 46.50 -- 47.40	\\
W3 J		& 48.16	& O8.5V	& IRS N3	& B0V -- B2V	& 46.50 -- 47.40	\\
W3 K		& 48.45	& O7.5V	& IRS N4	& B0.5V -- B2V	& $\leq$ 47.00	
\enddata
\end{deluxetable*}

\subsection{Energy budget of the \HII\ regions}

Apart from IRS N6, NR559 and NR386, all OB stars in W3 Main are located in \HII\ regions detected at radio wavelengths \citep{Tieftrunk97}. 
Assuming  the radio emission to be optically thin, we recalculated the number of ionizing photons responsible for the radio flux using the distance of 1.95 kpc instead of 2.3 kpc quoted by \citet{Tieftrunk97}. With \citet{Martins05} and \citet{Smith02}, the number of ionizing photons  are  converted in a spectral type of a single star that would be responsible for the ionization of the \HII\ region (Table \ref{tab:hii}). These spectral types are compared to the spectral types of the OB stars inside the \HII\ regions. In the case of W3 A (radio spectral type O6V), 5 OB stars are found in a single \HII\ region. Nevertheless,  IRS2, the earliest type star, is the main source of ionization as the number of ionizing photons drops very fast with spectral type \citep{Martins05}. IRS2 is of spectral type O6.5V -- O7.5V emitting 10$^{48.44 - 48.80}$ Lyman continuum  photons, while the 2nd most massive star, IRS2a (O8V -- O9V) only emits 10$^{47.90 - 48.28}$ ionizing photons.

For the \HII\ regions W3 A (O6.5V), W3 B (O7.5V), W3 C (O9.5V), W3 E (B1V) and W3 F (B0V), the most luminous stars provide sufficient Lyman continuum flux to ionize these \HII\ regions.  For the diffuse \HII\ regions W3 H (O9V), W3 J (O8.5V) and W3 K (O7.5V), on the other hand, the Lyman continuum flux of the  brightest source is not sufficient to explain the radio flux.  IRS N3 (B0V -- B2V) in W3 J and IRS N4 (B0.5V -- B2V) in W3 K are the brightest stars, located in the center of these \HII\ regions, and therefore are expected to be the main ionizing sources. However, the expected spectral type from the radio flux is much earlier and inconsistent with the observed spectrum of the ionizing sources. Interestingly, the photometric spectral type of IRS N4 (O7V) agrees better with that expected from the radio flux of W3 K. However, as discussed in Sect. 3.2, a O7V spectrum is not compatible with the the observed spectrum of IRS N4, so a miss classification of the spectrum is not likely.  The spectroscopic classification of IRS N3 and IRS N4 by \citet{Navarete11} confirms our spectral classification based on a different data set.

Leakage of FUV photons or dust attenuation would be able to explain why the ionizing star would be of earlier spectral type \citep[e.g.][]{Kurtz94}, but will not result in the  fact that the observed radio emission cannot be accounted for by the ionizing star. 
\citet{Smith02} show that the number of ionizing photons is much higher (log $Q_{\rm{0}}$ = 47.8 s$^{-1}$)  for a B1 super giant instead of log $Q_{\rm{0}}$ = 46.5 s$^{-1}$ for a B1 dwarf. This would come closer to the value derived from the radio emission. However, their position in the HRD (Fig \ref{fig:hrd}) does not comply with this explanation, since these stars are both too faint to be early B supergiants, which would have a luminosity of $\sim$10$^{5.5} L_{\sun}$ \citep{Martins05}, while the observed luminosity is $\sim$ 10$^{4.2} L_{\sun}$  (IRS N3) and $\sim$  10$^{4.5} L_{\sun}$ for IRS N4. The presence of supergiants is also not consistent with the derived age of W3 Main.

\subsection{Cluster mass}

Determinations of the  mass of embedded regions like W3 Main are hampered by the extreme extinction variations observed in the cluster  (Fig. \ref{fig:cmd}). No clearly reddened main sequence is visible which can be compared to main sequence isochrones directly \citep[e.g.][]{Ascenso07,Brandner08}.  However, the spectroscopy of the massive stars results in a direct mass estimate for the stars of the upper end of the mass function.

The masses of the stars were calculated by comparing the stellar luminosities to the 1 Myr main sequence isochrone. Only for IRS2, the 3 Myr isochrone is used as it already evolved away from the location of the 1 Myr isochrone. This results in stellar masses of 35 $M_{\sun}$ for IRS2 as the most massive star and 5 $M_{\sun}$ for IRS N6, the lowest mass star in our sample. 

In order to obtain an extinction-limited sample, which is complete  to the highest extincted massive star we have detected with spectroscopy, we took the mass of our reddest OB star, IRS 7 (11 $M_{\sun}$) as our lower mass limit where we are complete. In this way, we select 13 out of 15 OB stars studied in this paper. Inspecting the CMD shows  an additional 12 sources  above this mass limit without spectroscopic classification.  \citet{Navarete11} identified one of the bright stars, not related to any \HII\ region, as a massive star, this suggests that  the other stars might well be  members of W3 Main. Assuming all these 12 sources are part of W3 Main, we end up with a total of 25 stars between 11 and 35 $M_{\sun}$. This number does not include the  deeply embedded stars around IRS5 as they have even higher extinction. Additionally, our observations only cover the central 2.5 pc $\times$ 2.5 pc, while the cluster extends to 6 parsec based on X-ray observations \citep{Feigelson08}. However, the \HII\ regions tracing the massive stars are all located in the central 2.5 pc. We calculate the mass of the cluster assuming that we are complete in the detection of the massive stars with $A_{\rm{K_s}} \leq$ 5.9 mag.

Assuming a Kroupa IMF \citep{Kroupa02} with  a slope $\Gamma$ = -1.3 between 0.5 and 120 $M_{\sun}$ and -0.3 for masses between 0.08 and 0.5 $M_{\sun}$ and the total number of stars between 11 and 35 $M_{\sun}$  (25 stars), we can extrapolate the mass function and obtain an estimate of the total stellar mass in the cluster. For the extrapolation we carried out Monte Carlo simulations assuming a randomly populated IMF as described in \citet{Brandner08}, resulting in a total mass of  (4 $\pm$ 1) $\times$ 10$^3 M_{\sun}$.
 
 The quoted uncertainty is only due to the effects of random sampling. When only the stars where spectroscopy has confirmed their cluster membership are used (15 stars), then the derived mass is about a factor of two lower.  Additionally, binaries are not taken into account as well as the deeply embedded population like IRS5. This would make the real mass likely higher. With 4000 $M_{\sun}$, W3 Main is about twice as massive as the Orion  Nebula Cluster \citep{Odell01}.

The total mass of the W3 GMC is estimated to be 5$\times 10^{4} M_{\sun}$ based on $^{13}$CO observations \citep{Dickel80}, but encompasses more than just W3 Main alone. When looking at just the molecular gas towards W3 Main, \citet{Dickel80} derive a minimum mass for the high-density gas towards the W3 Main region of 5 $\times 10^{3} M_{\sun}$, which is confirmed by $^{12}$C$^{18}$O observations \citep{Thronson86}. Together with our mass estimate of $4 \times 10^3 M_{\sun}$, this would result in a estimate of the  star-formation-efficiency (SFE) of $\sim$44 \%.  Note that the southern part of the W3 Main region most likely already evacuated the gas, and therefore this should be considered as a rough upper limit.  Such high value of the SFE is consistent with the prediction of \citet{Bonnell11} for clusters forming in the gravitationally bound regions of a molecular cloud (SFE $\sim$ 40 \%). Whether the cluster will remain bound if the gas is removed \citep{Portegies10} remains uncertain and cannot be constrained with our current observations. 

\section{Discussion}

\subsection{massive Young Stellar Objects}

Apart from the 15 OB stars, 3 massive YSOs have been detected. Their spectra are dominated by circumstellar emission and no spectral type can be derived from their  $K$-band spectra. However, based on their location and detection at other wavelengths, we can derive some upper limits to their luminosity. IRS5 is located in a HC\HII\ region and is a small multiple system \citep{Megeath05,vanderTak05,Rodon08} with a total luminosity of 3 $\times$ 10$^5 L_{\sun}$ ionized by a handful of early B stars \citep{Campbell95}.  The two other massive YSOs in our sample, IRS N7 and IRS N8 are not associated with a detectable \HII\ region. Also, they have no detectable sub-mm counter part. IRS N7 is located at the southern edge of the sub-mm clump centered on IRS5, while IRS N8 is located east of the clump centered on IRS4 \citep{Moore07}.  

In contrast to the very embedded  IRS5, both sources are detected in $JHK_{\rm{s}}$ and therefore less embedded and probably more evolved than IRS5. The lack of radio emission would suggest that they are mid- or late B stars, as for more massive stars, an \HII\ region would be expected. Similar objects have been detected in other star forming regions \citep{Hanson97,Coletter04, Blum04, Brgspec06}, but no clear reason has been identified for why these stars are still surrounded by circumstellar matter, while other OB stars already cleared their surroundings. \citet{Brgspec06} showed that these objects have similar spectral and photometric properties  to  Herbig AeBe stars, which are still surrounded by circumstellar disks even when they become optically visible. 

We detected OB stars in the diffuse, compact and UC\HII\ regions. Only in the HC\HII\ region W3 M,  circumstellar emission from IRS5 has been detected in the $K$-band spectrum.  This shows that the accretion disks around the massive stars in W3 Main are destroyed within 10$^5$ yrs. When an HC\HII\ region evolves to the UC\HII\ phase,  circumstellar  disks are  destroyed by the strong FUV field and stellar winds of the recently formed massive stars \citep{Hollenbach94}.

\subsection{Evolutionary sequence in W3 Main}

W3 Main harbors several different evolutionary stages of \HII\ regions, ranging from very young HC\HII\ regions (few 10$^3$ yrs), UC\HII\ regions \citep[$\sim 10^5$ yrs][]{WoodIRAS89} to evolved,  diffuse \HII\ regions (few 10$^6$ yrs). All these regions are most likely formed out of the  same molecular cloud. This provides the possibility to study the evolution of young \HII\ regions and their stellar content in great detail.  \citet{Tieftrunk97} derived an evolutionary sequence for the \HII\ regions in W3 Main based on the morphology of the radio sources.  
The youngest are the HC\HII\ regions W3 M and W3 Ca, with the UC\HII\ regions W3 F, W3 C and W3 E, slightly older, the compact \HII\ regions W3 B and W3 A even more evolved, and the diffuse \HII\ regions, W3 K and W3 J being the oldest \HII\ regions in W3 Main. This classification can be compared to the ages of the massive stars deduced from their position in the HRD. 

We have detected OB stars in three diffuse \HII\ regions, two compact \HII\ regions and three UC\HII\ regions with  HC\HII\ region W3 M harboring the high-mass protostar IRS5 (see Sect. 4.1).  The position of IRS2 in the HRD suggests an age of  2-3 Myr, consistent with its location in a relatively evolved compact \HII\ region (W3 A) with a similar or younger age derived for  IRS3a,  harbored in a younger UC\HII\ region. For the lower-mass stars (late O, early B) the isochrones are too close to each other to derive any age information. For IRS N1, located inside the UC\HII\ region W3 E, the offset from the main sequence could be explained by high-mass PMS evolution \citep{Hosokawa09}, consistent with the expected young age of the UC\HII\ region.
 
A similar sequence can be seen in the extinction towards the different \HII\ regions. The extinction varies from $A_{\rm{K_s}}$ = 0.9 mag for the diffuse \HII\ regions W3 J and W3 K to very high extinction ($A_{\rm{K_s}}$  = 5.9 mag) towards  W3 F. Only IRS3a, located in the compact \HII\ region W3 B does not follow the trend and shows a very high extinction (A$_{\rm{K_s}}$  = 5.4 mag). Such a sequence in extinction indicates that the stars in  diffuse \HII\ regions have already cleared out their surroundings and destroyed the molecular cloud, while those in UC\HII\ regions are still  embedded in their parental molecular cloud.  

Based on the presence of different evolutionary stages of \HII\ regions as well as the location of the most massive stars in the HRD, we can conclude that an age spread of a few Myr is most likely present for the massive stars in W3 Main. The presence of an older population of a few Myr is supported by the detection of several 100s class III stars in the X-rays  by \citet{Feigelson08}. 
These authors suggest that, at least some, of the OB stars have formed after the lower-mass PMS stars. A spectroscopic analysis of a sample of intermediate-mass PMS stars will provide more insights on this suggestion (Bik et al, in prep).

A large number of young stellar clusters show evidence for an age spread, usually based on the analysis of the PMS population in the HRD. In Orion an age spread of a few Myr has been found  \citep{Palla99,daRioOrion10},  as well as in the LMC in the young stellar cluster LH95 \citep{daRioLH9510}. In starburst clusters, however, upper limits on the age spread of less than 1 Myr have been found for Westerlund 1 \citep{Clark05,Negueruela10} based on the massive stars. Similar results are found for the PMS population of NGC 3603 and Westerlund 1 (Kudryavtseva et al, in prep). This suggests that W3 Main, similar in mass to the Orion Nebula cluster,  is not formed in one star formation burst as expected for starburst cluster, but, more likely, through a temporal sequence of star formation events. 

\section{Conclusions}

We have obtained deep $JHK_{\rm{s}}$ near-infrared imaging as well as $K$-band multi-object spectroscopy of the massive stellar content of W3 Main using LUCI at the LBT.   We have confirmed the nature of 15 candidate OB stars by means  of their absorption line spectra in the K-band.  Three additional sources are identified as massive Young Stellar Objects based on their emission line spectra.

From the  analysis of the CCD and CMD, the extinction laws of \citetalias{Indebetouw05} and \citetalias{Nishiyama09} provide the best fit of the slope of the reddened main sequence.   Using either of the these two extinction laws the OB stars are placed in the HRD,  with the \citetalias{Nishiyama09}  providing a better match between the derived positions of the stars and the main sequence. 

Our most massive star, IRS2 ($35~M_{\sun}$) is already evolved away from the ZAMS and its location is consistent with an age of 2-3 Myr.  Additionally, we find evidence for high-mass PMS evolution in the displacement of IRS N1 and IRS N4 from the main sequence. A total cluster mass of ($4 \pm 1) \times 10^3~M_{\sun}$ has been derived by extrapolating the number of massive stars using a Kroupa IMF.
 
The evolutionary sequence of the \HII\ regions seen in the radio morphology is consistent with a increase in extinction from the older (diffuse \HII) regions to the younger (UC\HII) regions.  We have detected the photospheres of OB stars from the more evolved diffuse \HII\ region to the much younger UC\HII\ regions, suggesting that the OB stars have finished their formation and cleared away their possible circumstellar disks very fast. Only in the HC\HII\ phase (IRS 5), the massive star is still surrounded by circumstellar material. 

The above properties of the stellar content and \HII\ regions present evidence of an age spread of at least 2-3 Myr in W3 Main, similar to what has been observed in Orion and other young open clusters and in contrast to the single age population found in starburst clusters. This suggests  a sequential star formation mode being at work for W3 Main, instead of a single star formation event.

\acknowledgments
We thank the anonymous referee for useful comments, helping to improve the paper. We thank the LBTO staff for their support during the observations. We thank Jaron Kurk and  Neil Crighton for obtaining the long slit spectra of IRS4 and IRS7. We thank Giovanni Cresci and Jaron Kurk for providing the lucired pipeline and  Takashi Hosokawa and Rolf Kuiper for the high-mass PMS tracks. A.B thanks Angela Adamo for reading the manuscript. D.A.G. kindly acknowledges the German Aerospace Center (DLR) and the German Federal Ministry for Economics and 
Technology (BMWi) for their support through grant 50~OR~0908.

{\it Facilities:} \facility{LBT(LUCI I)}.


\end{document}